# Simplification of Multi-Scale Geometry using Adaptive Curvature Fields

Patrick Seemann[1], Simon Fuhrmann[2], Stefan Guthe[1], Fabian Langguth[1], and Michael Goesele[1]

[1] GCC, TU Darmstadt, Germany [2] Google Inc.

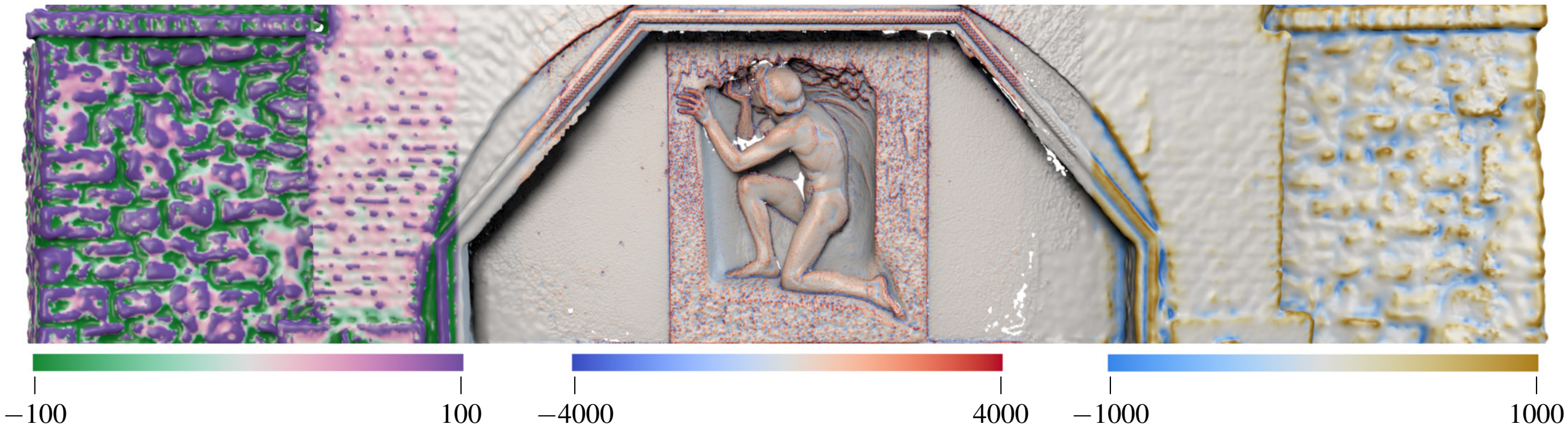

Figure 1: A single mean curvature field visualized on the mesh surface. In multi-scale meshes, the scale between the curvature values vary by several orders of magnitude. This is illustrated here using three colormaps that show the curvature field at different scales.

**Abstract**

*We present a novel algorithm to compute multi-scale curvature fields on triangle meshes. Our algorithm is based on finding robust mean curvatures using the ball neighborhood, where the radius of a ball corresponds to the scale of the features. The essential problem is to find a good radius for each ball to obtain a reliable curvature estimation. We propose an algorithm that finds suitable radii in an automatic way. In particular, our algorithm is applicable to meshes produced by image-based reconstruction systems. These meshes often contain geometric features at various scales, for example if certain regions have been captured in greater detail. We also show how such a multi-scale curvature field can be converted to a density field and used to guide applications like mesh simplification.*

## 1. Introduction

Triangle meshes are the most common geometry representation and their properties have been studied extensively in order to visualize, analyze, and modify them effectively. An important geometric property is surface curvature. However, due to the discrete nature of triangle meshes the computation of their curvature values is non-trivial. In practical scenarios noise can have a strong influence on the output of estimation algorithms. To cope with these problems recent techniques [YLHP06, SHBK10, APM15] apply a smoothing operator which successfully removes noise but ultimately also affects the geometric detail. The biggest problem is to select an appropriate scale for this operator. If the scale is chosen too small, the noise will interfere with curvature estimation; if the scale is chosen too large, surface details will be smoothed away. This problem is even more pronounced for multi-scale geometry. Image-based geometry acquisition pipelines using multi-view stereo (e.g., [FLG14]) can generate surfaces on vastly different scales depending on the camera resolution and its distance to the real-world objects as illustrated in Figure 2. The resulting triangle meshes then contain geometric features and noise on different levels of detail. A single scale curvature estimation cannot capture the true properties of the whole surface.

In this work we present a novel algorithm that estimates the curvature field of multi-scale triangle meshes. Previous methods [YLHP06, SHBK10] compute curvatures by evaluating a neighborhood around a given vertex using the ball neighborhood, which we also use in our work. Integral invariants (Section 3.1) can then be used to compute the mean curvature using the neighborhood of a vertex within the ball radius. The chosen radius defines the scale at which features are preserved and noise is smoothed. If the radius of the ball is fixed, the operator uses a uniform scale and cannot adapt to the scale variations of the surface. As a result the operator smoothes too much detail or retains too much noise.

Our main contribution is the independent and automatic selection of an appropriate ball radius for each vertex. Our method is robust against large variations in scale and can effectively distinguish between noise and geometric features. It operates directly on the mesh representation and does not require a volumetric shape representation. It is able to handle difficult input data, such as the meshes produced by image-based reconstruction techniques, which usually have varying level of detail and contain many holes.

A direct application of our method is mesh simplification. Particularly in image-based reconstruction scenarios the resulting meshes often contain millions of triangles because the vertex sampling is determined by the resolution of input images, not the geometric properties of the surface. We show how our estimated curvature field can be effectively transformed into a density field that guides the simplification process. As a result, the simplification algorithm does not need to be concerned with preserving geometric features of the surface. Instead, its task reduces to producing a vertex distribution prescribed by the density field, thus preserving more geometric detail in regions of higher curvature.

## 2. Related Work

Curvature estimation on discrete surfaces has been thoroughly studied and can be classified into local fitting methods, methods based on the angles between edges, and integral invariant-based methods, which integrate over larger surface regions. The latter methods are most promising in our scenario and usually perform better if the meshes are large, have geometric features at various scales, varying level of detail and noise. Many methods estimate curvature on a user-provided scale and compute curvature using a neighborhood with fixed radius. These methods do not perform well on multi-scale geometry because a suitable radius does not exist. Multi-scale methods, on the other hand, try to determine a suitable radius for each vertex.

### 2.1. Curvature Estimation on a Fixed Scale

There are many algorithms for computing curvature on a fixed scale [YZ13, ASWL11, MOG09]. The scale is usually provided by the user as input. Seibert et al. [SHBK10] make use of geometric algebra and compute principal curvatures directly on point set surfaces. Their approach estimates curvature at each vertex $x$ by fitting osculating circles in uniformly sampled directions around $x$ to a fixed local neighborhood of points. The principal curvatures for each vertex are obtained by combining the radii of the osculating circles for all directions. Because their approach relies on least squares fitting, a dense vertex sampling is required, and noise and outliers quickly degrade the curvature estimation. Their algorithm is not applicable to multi-scale geometry because it operates on a fixed local neighborhood around each vertex.

Andreadis et al. [APM15] compute geometric features (such as mean curvature [PWY*07]) at multiple scales by first transforming the input mesh into a parametric space. This decouples the computational complexity from the underlying geometry in order to produce a GPU-friendly, highly performant algorithm. However, their method relies on a mesh parameterization which has to be precomputed and is more difficult for less controlled meshes (higher genus, boundaries and holes in the surface, etc). The scale at which the curvature is computed is fixed and provided as user input.

Other approaches are based on integral invariants. Yang et al. [YLHP06, PWY*07] were the first to use integral invariants for robust estimation of curvature information of 3D meshes. To this end, the authors define the ball, sphere and surface-patch neighborhoods and perform a principal component analysis (PCA) on each neighborhood. They derive formulas to calculate the two principal curvatures and the mean curvature based on volume integral invariants from the PCA. Their definition yields the notation of curvatures at a scale $r$, where $r$ corresponds to the radius of the neighborhood. The authors claim that their approach is more robust than normal cycles [CSM03] and local fitting methods like osculating jets [CP05]. In particular, the ball neighborhood seems suitable for noisy input data. Our approach is based on the ball neighborhood, and we extend their method by robust and automatic, per-vertex scale selection over a large range of scales.

### 2.2. Multi-Scale Curvature Estimation

Multi-scale algorithms try to choose an appropriate scale for each vertex at which the curvature is estimated. Usually, the user specifies a lower and upper bound instead of a single scale. Lai et al. [LHF09] also use integral invariants based on the ball neighborhood to compute multi-scale principal curvatures. Instead of relying on user input to specify the scale of interest, an iterative algorithm adjusts the radius $r$ of the ball neighborhood for each vertex independently. A series of $n$ subsequent ball radii between a lower and upper are evaluated, and principal curvatures are computed for each radius. The algorithm uses the pre-computed curvatures and interpolates new, refined radii until convergence.

Choosing the lower and upper bound, however, remains a challenging problem. The authors propose to determine these values as factors of the average edge length in the mesh. While this solution works in the authors scenario where evaluation is performed on meshes with almost identical triangle sizes, we target true multi-scale meshes where the triangle sizes vary substantially. Thus, a global starting radius cannot be defined, this is illustrated in Figure 2. Our method improves this aspect and selects the per-vertex radius using the edge lengths in the local neighborhood of a vertex.

Another drawback of their method is that it requires a voxel representation of the model to approximate the volume of the ball neighborhood. However, a voxelization (e.g., using scan conversion) causes problems when the input mesh is not closed or has many holes. Our algorithm computes the ball neighborhood at a given vertex without relying on a volumetric representation.

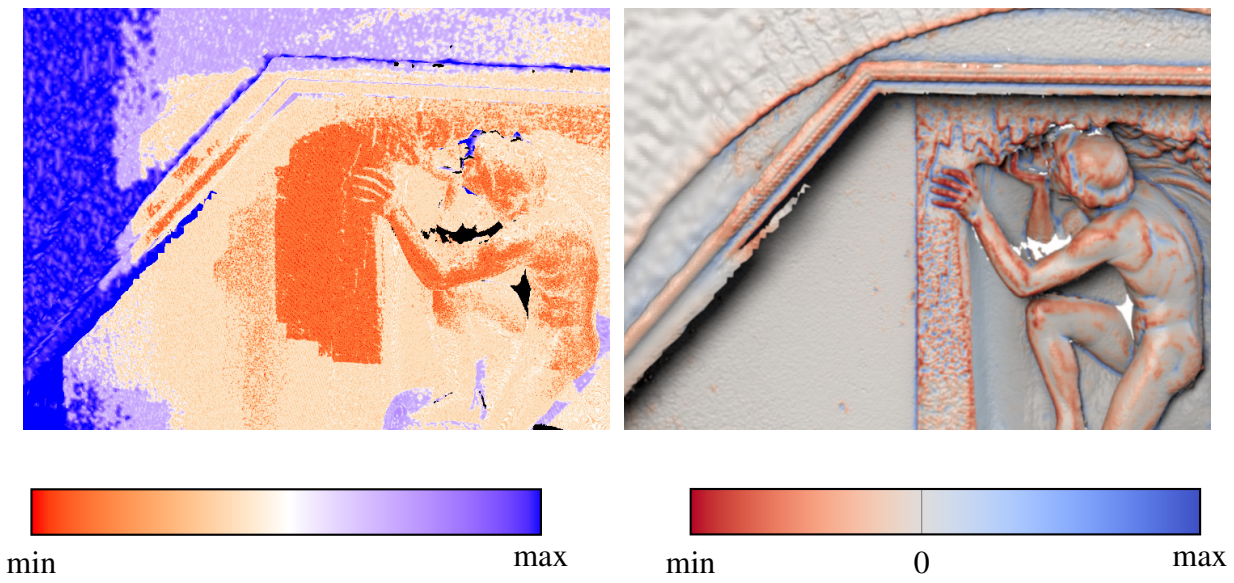

Figure 2: The average edge length in a 1-ring around a vertex (left) varies drastically throughout the mesh. Our mean curvature field (right) is not influenced by the different triangle sizes and produces correct values. To increase readability, the average edge length was clamped to the $1^{\text{th}}$- and $90^{\text{th}}$-percentiles and plotted in log-scale. In this mesh, the smallest triangles are about 42 times smaller than the largest ones.

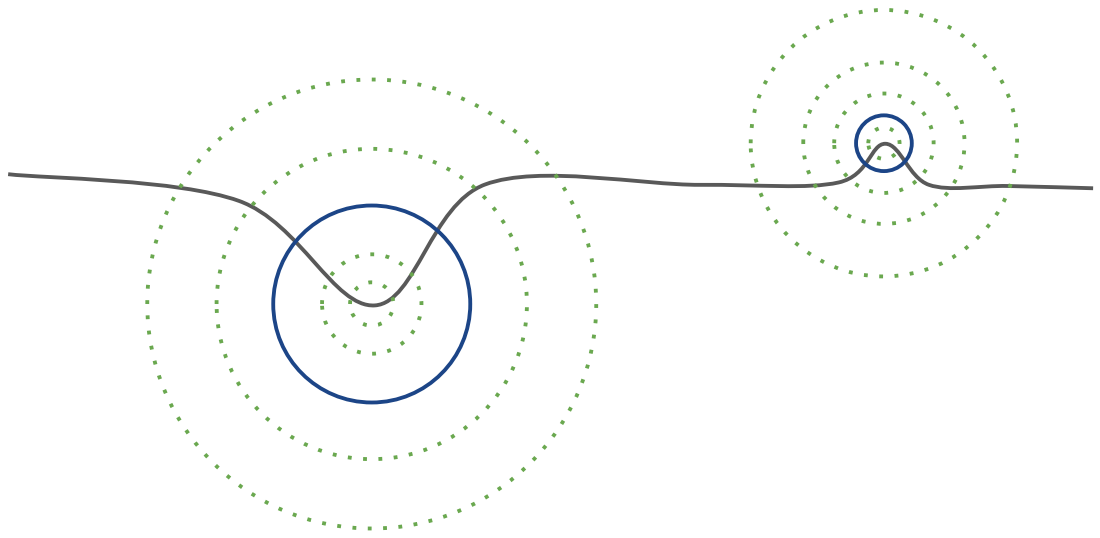

Figure 3: For each vertex, we compute the mean curvature at different radii (green circles) and then select the correct radius for each vertex (blue circles).

## 3. Algorithm

Curvature is defined as the second-order derivative of the mesh surface and thus inherently depends on its scale. Given the discrete nature of a mesh, it must be evaluated numerically over an appropriately sized neighborhood. Additionally, if the mesh contains noise, a large enough neighborhood must be found that cancels out the noise while maintaining important surface details.

The first and most involved step of our algorithm computes the curvature field on the mesh surface. For each vertex of the mesh, a ball with an appropriate radius is found and used to compute a signed mean curvature value using integral invariants. We then motivate how the curvature field can be used for the purpose of mesh simplification by converting it to a density field. The density field prescribes the relative vertex sampling density to faithfully represent the mesh surface at a given vertex budget, by distributing more vertices in regions of higher curvature. As a result, the simplification algorithm is not concerned with preserving geometric features during decimation. Instead, it merely selects vertices for decimation which have a small amount of density associated with it.

### 3.1. Integral Invariants

A detailed analysis of integral invariants was published by Manay et al. [MHYS04]. In essence, they are used to compute integral quantities (as opposed to differential ones) over different types of neighborhoods. The invariance depends on the function that is used to compute it. In our case, we compute the volume integral, which yields the invariance to mesh rotation and translation. Because one does not have to compute higher order derivatives, integral invariants are more stable in the presence of noise. The neighborhood on triangle meshes is defined by the surface of the mesh and the volume which it represents. Pottmann et al. [PWY*07] propose the sphere, ball, and the surface-patch neighborhoods. They analyze these neighborhoods with respect to their noise properties and conclude that the ball neighborhood performs best. We use this neighborhood when computing our volume integral invariants.

The formal definition of the volume integral invariant using the ball neighborhood is as follows. Let $D$ be a domain and $\Phi$ its boundary surface. For any point $p \in \Phi$ and ball $B(r,p)$ with radius $r$ centered at $p$ one can compute the intersection volume for the neighborhood around $r$ as $D \cap B(r,p)$. We will refer to the magnitude of this volume as $V_b(r,p)$. Pottmann et al. [PWY*07] derive a formula for computing the mean curvature $H$ from the ball neighborhood at a given radius $r$:

$$H_{ball}(r,p) = \frac{4}{\pi r^4}\left(\frac{2\pi}{3}r^3 - V_b(r,p)\right) \quad (1)$$

This discrete mean curvature converges to the actual continuous mean curvature for $r \to 0$. For larger radii, the mean curvature is smoothed. As Lai et al. [LHF09] noted, the radius $r$ is of special importance since larger radii produce smoother results which are thus more robust to noise. On the other hand, there is a risk of smoothing away small geometric features when the radius is too large. Therefore, a single radius is not applicable for meshes with multiple scales. Because in some regions the radius will be too large and smoothes away important surface details. In other regions, the radius will be too small and produces an unwanted response in the presence of noise.

### 3.2. Curvature Field Computation

To compute the curvature field of a triangle mesh, we first evaluate the mean curvature at each vertex using Equation 1 on multiple radii (see Section 3.5) and eventually select the correct radius for each vertex, see Figure 3. For the calculation of the integral invariant, the volume of the ball neighborhood has to be evaluated for each individual radius. We approximate this volume using a triangulated sphere with an adaptive tessellation along the intersection border (see Section 3.4). To intersect the sphere with the surface of the mesh, a circular surface patch consisting of all faces within the the current radius (see Section 3.3) is used. For finding the final radius, we fit a cubic polynomial to the collected data as described in Section 3.5.

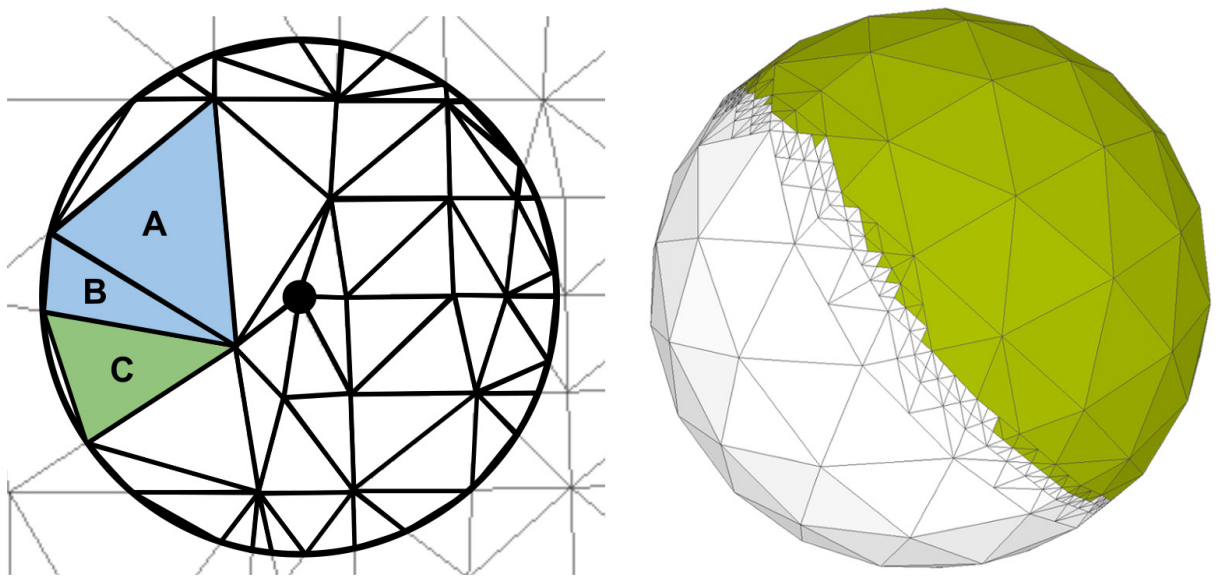

Figure 4: Triangulation of a circular patch (left): New triangles are created where the circle intersects a face. Two triangles (A and B) are created if one vertex is outside the radius. One triangle (C) is created if two vertices are outside. Faces fully within the radius are simply copied into the patch. The right picture shows the adaptive intersection (green faces) of a sphere with the surface.

### 3.3. Surface Patch

For calculating the volume of the ball neighborhood $V_b(v_c, r)$ around a vertex $v_c$, we first find the surface patch contained within the radius $r$ using a region growing approach. We start by iterating over all faces $f_i$ in a 1-ring around $v_c$. For each vertex $v_j$ in $f_i$ we check whether it is inside the radius by calculating the Euclidean distance to the center of the sphere: $\|v_j - v_c\| \le r$. We then add every vertex where this inequality holds to a list of vertices, which we visit in the next iteration. If all three vertices of the face $f_i$ are within the radius, we simply add $f_i$ to the patch. If some vertices are outside the radius, we cut the face $f_i$ at the intersection points of the radius with the edges of the face and create new triangles with the intersection points. There can only be two cases: Either one or two vertices of $f_i$ are outside the radius. See Figure 4 for an illustration. When the algorithm terminates, the result is a circular surface patch. As a final step, we calculate the face and vertex normals of all triangles in the patch. These will later be used to compute the intersection volume.

### 3.4. Volume Computation using a Triangulated Sphere

The volume of the ball neighborhood is approximated by intersecting a triangulated sphere with the triangulated shape. The surface is represented by the surface patch that was computed in the previous step. Using the triangles of the patch as well as all faces of the sphere which are behind the surface, the volume of the resulting polyhedron can be computed by the following formula:

$$V_{approx.} = \frac{1}{6} \sum_{i=0}^{N-1} a_i \cdot \hat{n}_i \tag{2}$$

where $N$ is the number of triangles and $\hat{n}_i$ is the unnormalized normal of triangle $f_i = (a_i, b_i, c_i)$ calculated by $\hat{n}_i = (b_i - a_i) \times (c_i - a_i)$.

For numerical stability, we first translate the surface patch to the origin. Thus, the sphere is also centered at the origin and scaled by the current radius. Computing the actual intersection of the sphere and the patch is non-trivial and costly. We therefore approximate the intersection by finding all faces of the sphere which lie behind the surface. We check whether each vertex of the sphere is behind or in front of the surface by finding its nearest neighbor vertex on the surface patch and perform an inside/outside check based on the scalar product of both vertex normals. We accellerate this nearest neighbor lookup, using a $k$-D tree data structure with all patch vertices. To compute the final volume, we use Equation 2 and sum over all patch faces as well as all faces of the sphere behind the mesh surface.

The triangulated sphere is generated using two Loop subdivision iterations [Loo87] on an icosahedron, which result in 162 faces. The sphere faces along the intersection with the mesh surface are further subdivided as illustrated in Figure 4. Experiments have shown that six subdivisions along the border result in negligible approximation error for the intersection boundary.

### 3.5. Radius Sampling

To get a suitable initial radius for the current vertex $v$, we use the average edge lengths in a 1-ring neighborhood around $v$. Let $s(v)$ denote the scale of a vertex $v$. The starting scale $s_0$ of $v$ is then computed by

$$s_0(v) = \frac{\sum_{w \in N(v)} \|w - v\|}{|\, N(v) \,|} \tag{3}$$

where $N(v)$ is the set of all neighbors of $v$. This approach already produces a good starting radius for each vertex. In order to cope with regions where the edge lengths are extremely small or large, we smooth the initial radius afterwards. For this, we perform $n$ multiple smoothing iterations and update the initial scale of each vertex $v$ based on its surrounding vertices:

$$s_{i+1}(v) = s_i(v) + \sum_{w \in N(v)} \lambda \frac{s_i(w) - s_i(v)}{|\, N(v) \,|} \tag{4}$$

where $\lambda$ is a smoothing factor. A larger value of $\lambda$ increases the influence of neighboring vertices on $v$. If overall smoother results are desired, the starting radius can be multiplied with an additional factor $> 1$.

To sample the mean curvature at different radii $r_0, \ldots, r_n = s_0, \ldots, f_{max} \cdot s_0(v)$, we exponentially increase the radius by $f_{inc}$ until it reaches a predefined maximum $f_{max}$. In each iteration $i$, we compute the mean curvature for a vertex $v$ at radius $r_i = r_{i-1} \cdot f_{inc} = r_0 \cdot f_{inc}^i$. For all of our experiments we use $f_{inc} = 1.3$ and $f_{max} = 10$ (corresponding to $n = 8$) which results in a large enough sampling region to distinguish small features from noise while still being able to detect planar regions.

This sampling yields two dimensional data $D_v = \{(r_i, H(r_i, v)) \mid r_i \in \{r_0, ..., r_n\}\}$ for each vertex $v$. Because the sampling is discrete, we fit a function to $D_v$ in order to make a decision for the final radius in continuous space. This fit, however, must be performed on the normalized curvature to not be influenced by the smoothing introduced through the radius of the ball neighborhood. Using Equation 1 we get the normalized curvature:

$$H_{norm}(r, V_b) = r \cdot H_{ball}(r, p) = \frac{8}{3} - \frac{4}{\pi r^3} V_b. \tag{5}$$

Equivalently to the unnormalized case, $H_{norm}$ is zero when $V_b$

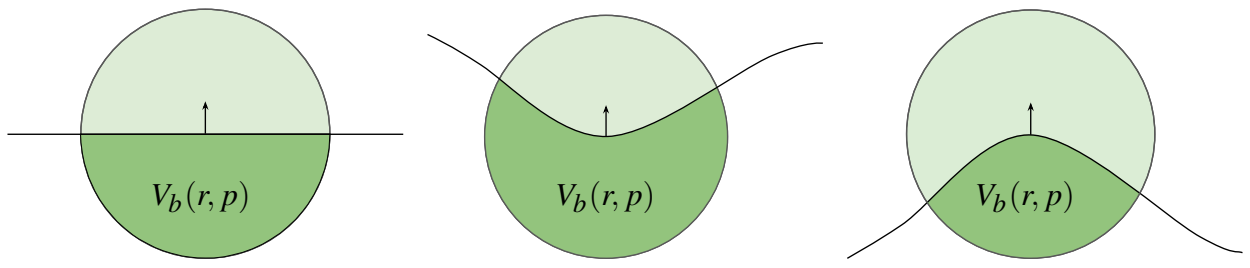

(a) $H_{norm} = 0.0$ (b) $H_{norm} < 0.0$ (c) $H_{norm} > 0.0$

Figure 5: The surface is either planar and has zero mean curvature or it is curved and thus has a negative or positive mean curvature. $V_b(r,p)$ is the volume of the ball neighborhood with radius $r$ at point $p$.

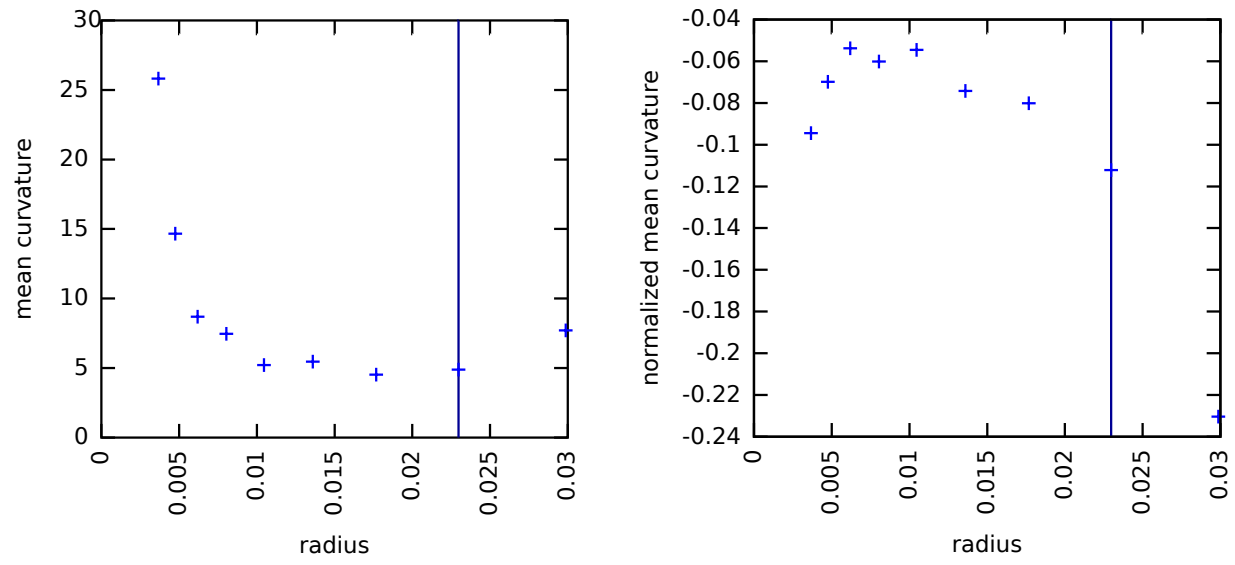

Figure 6: The final surface patch for four vertices which were classified to be in a planar region. The graphs show the mean curvature (left) and the normalized curvature (right) for a single vertex. Here, the curvature at radius 0.03 is treated as an outlier.

is exactly half of the ball volume; for larger and smaller values we get $H_{norm} < 0$ and $H_{norm} > 0$ respectively, see Figure 5 for an illustration. This leads to scale invariant, normalized data $\hat{D}_v = \{(r_i, H_{norm}(r_i, v)) | r_i \in \{r_0, ..., r_n\}\}$ for each radius increase. We fit a cubic function in the least-squares sense to the data $\hat{D}_v$ for each vertex. In our experiments a quadratic fit often does not represent the data well enough while higher order polynomials cause problems because of overfitting. Thus, we check if the error of the fit is small (below 2%). Otherwise we try to optimize it by iteratively removing data samples starting with the sample at the largest radius.

### 3.6. Radius Selection

The final radius for each vertex is chosen in different ways depending on the local surface properties. We first decide if the vertex lies on a planar or non-planar region. In the latter case, we analyze the extrema of the fitted polynomial which can have up to two extrema in the given interval. The following cases are considered.

**Planar regions:** If a surface region is planar, then the normalized curvature values are close or equal to zero. We use the average of the normalized curvatures of the current vertex and check whether it is below the planar threshold of $t_p = 0.2$, which is a good choice for most meshes. If the surface of the triangle mesh is particularly noisy, $t_p$ can be increased, which will result in more smoothing. Whenever a planar region is detected, we select the largest radius at which the normalized curvature is below the threshold, see Figure 6.

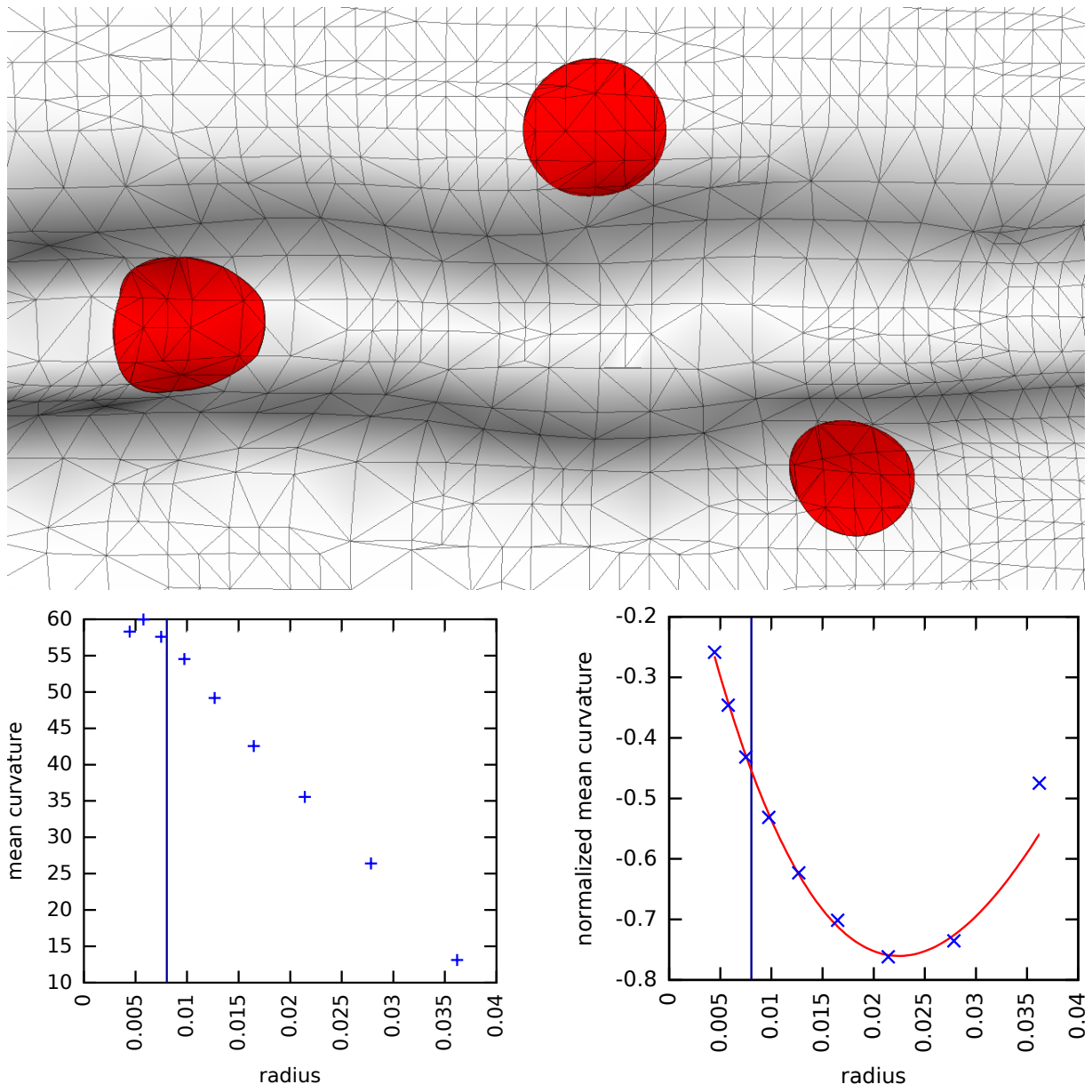

Figure 7: Example vertices around a valley which lie in a non-planar region. The extreme point of the fitted polynomial is used to select the final radius by interpolating between the smallest radius and the radius at the extreme point based on the edge smoothing factor $f_{es}$.

**Single extrema:** One extreme point within the sampled radius range is an indicator for a region which starts off with high curvature. Because the smoothing performed by the integral invariant based mean curvature computation, we cannot simply choose the radius at the extreme point as the final radius because the closer we choose a radius towards the extreme point, the smaller the mean curvature gets (Figure 7). We therefore introduced a global "edge smoothing" factor $f_{es}$. This factor is used to control the final radius in such a case where $f_{es} = 0.0$ results in the smallest radius $r_0$ and $f_{es} = 1.0$ corresponds to the radius at the extreme point.

**Two extrema:** This indicates geometric detail at the first extrema and another significant enough surface feature at the second. The first extrema is used to compute the final radius to preserve surface detail. We select this final radius by applying the edge smoothing factor as described in the previous paragraph.

**No extrema:** This can be observed when the polynomial either has no extreme points or when both extreme points are outside of the radius range. E.g., consider an edge where $H_{norm}$ converges towards $-\frac{4}{3}$ or $\frac{4}{3}$, for a 90°or 270°edge respectively. Note that $H_{norm}$ is not constant because edges in meshes resulting from surface reconstruction algorithms are typically not sharp in contrast to meshes used in Computer Aided Design. If there is a saddle point, we use it as a reference and interpolate between it and the smallest radius based on $f_{es}$. Otherwise we use the middle radius.

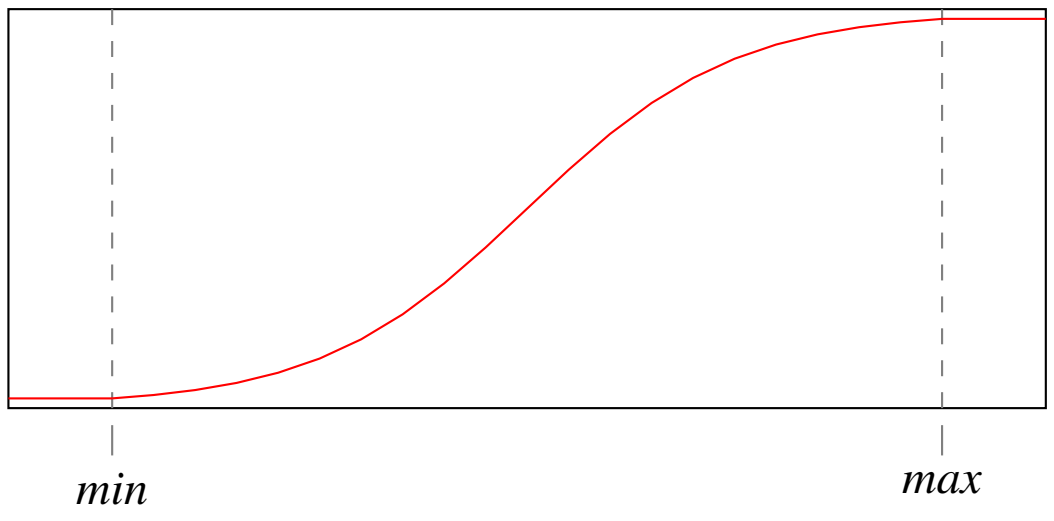

Figure 8: To create a density field useful for mesh simplification, the curvature values are remapped using the density function $d(x)$.

### 3.7. Density Field Computation

For mesh simplification the curvature field is remapped to a range more suitable for this application. We need to define the importance of a vertex as a single positive number. The sign of the curvature is not important, so we only consider the absolute mean curvature value. The curvature values are remapped (Figure 8) such that the density values are constant up to a minimum and then increased up to a maximum using a sigmoid function.

The exact definition of the mapping we used is as follows:

$$d(x) = \begin{cases} d_{min} & \text{for } x \leq min \\ d_{scale} \cdot \left( \frac{1}{1+e^{-4\hat{x}}} - \frac{1}{1+e^{4}} \right) + d_{min} & \text{for } min < x \leq max \\ d_{max} & \text{for } x > max \end{cases}$$

with $\hat{x} = 2\frac{x-min}{max-min} - 1$ and $d_{scale} = \frac{d_{max}-d_{min}}{\frac{1}{1+e^{-4}} - \frac{1}{1+e^{4}}}$.

$d_{min}$ and $d_{max}$ correspond to the minimum and maximum density values and should be chosen according to the mesh simplification algorithm that is used. A final smoothing on the density field helps to gradually change the triangle sizes from low to higher density regions. In our experiments, however, we found that this smoothing is not absolutely necessary and depends on the application.

## 4. Results

We evaluate our algorithm on several real-world datasets which were created using the open source image-based reconstruction pipeline of [FLM*15]. Mesh simplification was performed using the *Remesher* tool from [FAKG10] together with our density fields. The mesh properties and the runtimes of our implementation are summarized in Table 1.

In Figure 9 we show the *Bronze Akt* dataset. The statue has a very rough surface and the reconstruction produced different levels of geometric noise, resulting from pictures taken from various distances to the model. In order to extract curvatures in a meaningful way, the radius for the ball neighborhood must be chosen such that small scale features which can be found in the platform region are preserved while noise on the overall model is smoothed. Choosing a single radius that preserves small-scale features results in a noisy curvature estimate on the rest of the surface. Using a larger radius removes most of the noise but also smoothes away important features. Our algorithm preserves the features while also removing noise by adaptively varying the radius per vertex.

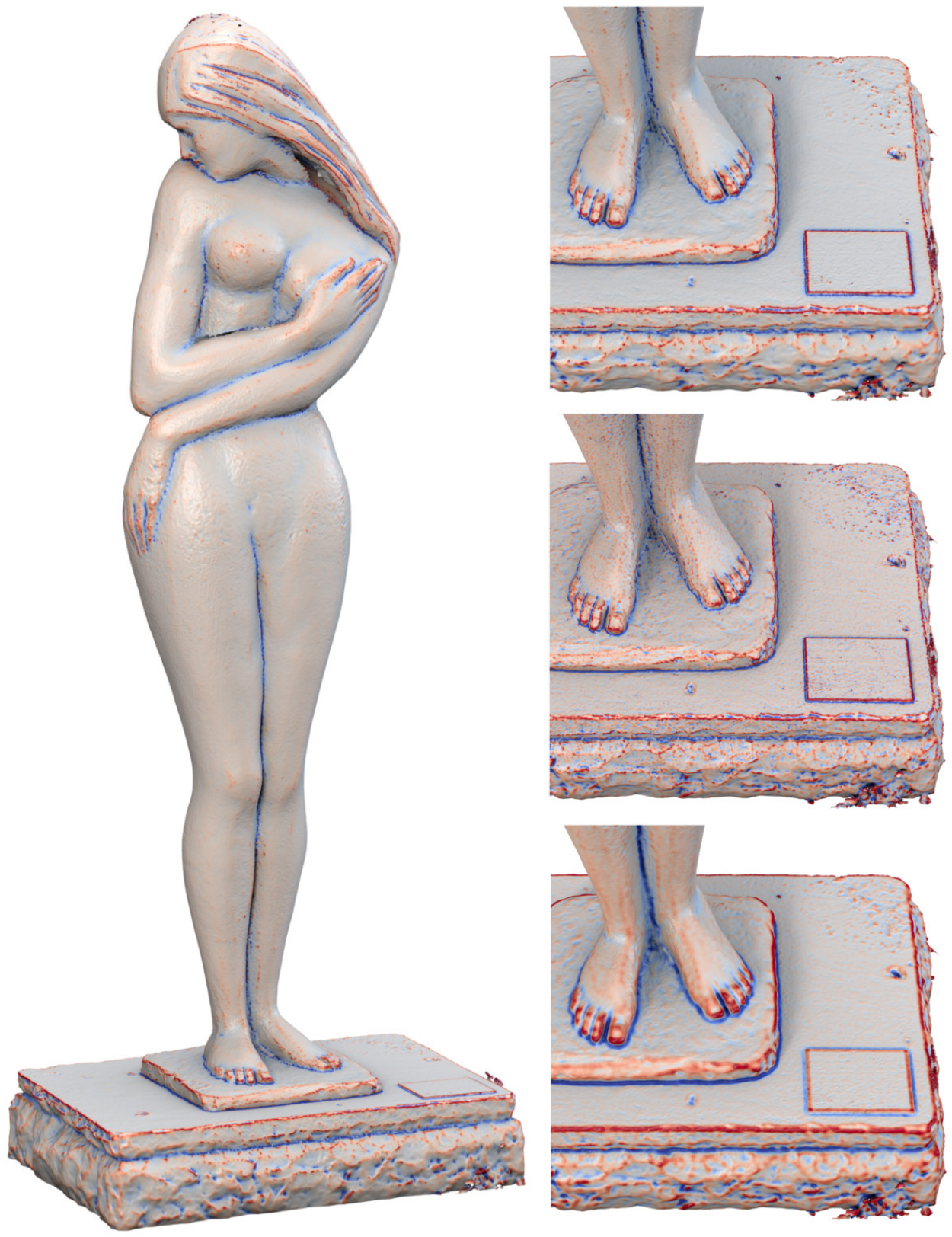

Figure 9: *Bronze Akt*: Our curvature field (left) as well as three closeups of the platform region. Our result (top), curvatures computed using a fixed small radius (middle) and fixed larger radius (bottom).

Simplification of multi-scale geometry is a direct application of using the density field described in Section 3.7. Figure 10 shows a model of the *Fountain* dataset, which has many small-scale features at the statue. We simplified this mesh to 4% of the original number of vertices to achieve a size that is manageable for real-time or mobile applications. Using a single-scale density field results in a direct loss of small-scale features. Our multi-scale density field guides the simplification to preserve important features such as the hand of the statue.

The effects of using a single-scale radius can be also be seen in the *Goethe-Fountain* dataset, Figure 11. This dataset is another example with significant scale differences. A globally selected radius cannot cover all of the geometry features that are captured in different resolutions. During simplification we reduced the amount of vertices to 3%. The details on the fountain head are lost or edges are smoothed too much when a single-scale curvature is used. Our multi-scale curvature estimate leads to a significantly better simplification result which still retains most of the geometric detail from the original mesh.

The *Owl* dataset does not contain scale differences. Here we show that our algorithm gracefully degenerates to estimating curvature at a single scale.

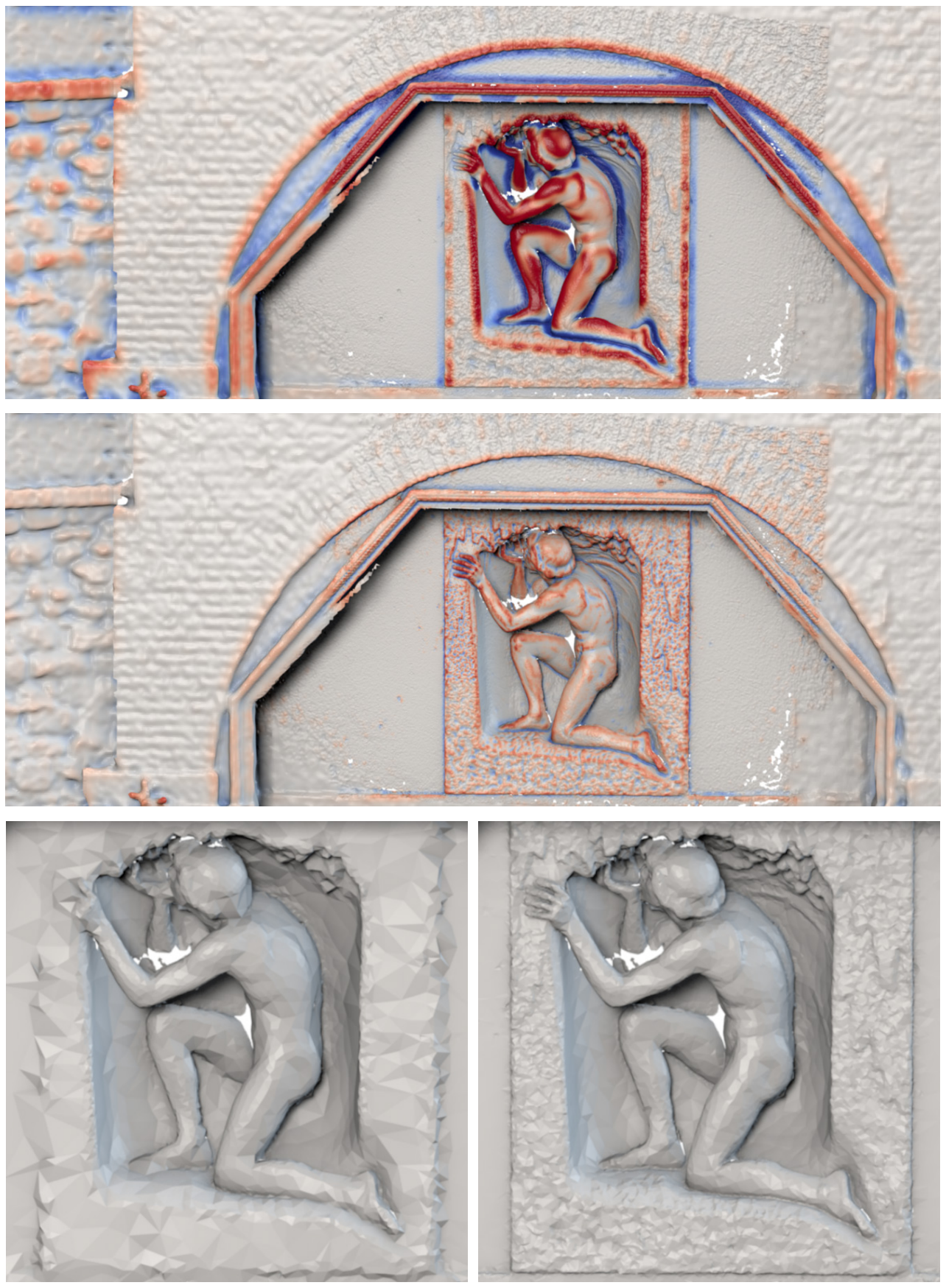

Figure 10: *Fountain*: Simplification without density field, single-scale curvature field (top), our multi-scale curvature field (middle), simplification using the single-scale curvature field (bottom left) and our simplification (bottom right). We simplified the original mesh to 4% in both cases.

| Dataset Name | # Vertices | Runtime |
|---|---|---|
| Owl | 280 752 | ~11 min |
| Fountain | 2 068 619 | ~65 min |
| Bronze Akt | 3 666 075 | ~111 min |
| Goethe-Fountain | 4 675 851 | ~159 min |

Table 1: Runtime of our algorithm for estimating multi-scale curvatures. The total computation time increases linearly with the number of vertices.

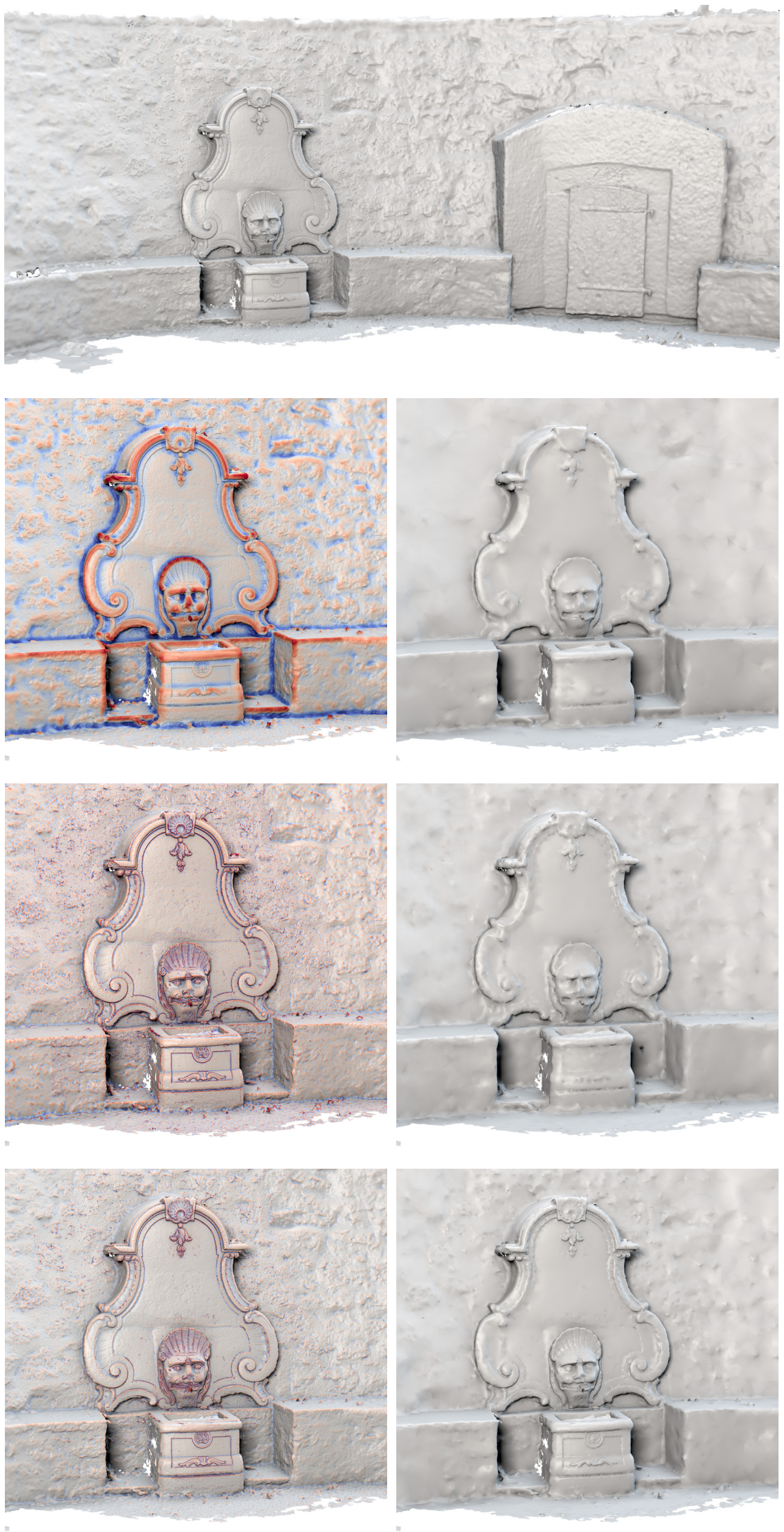

Figure 11: *Goethe-Fountain*: overview shot (top), curvature fields (left column) and corresponding simplifications (right column): single-scale curvature large radius (top) and small radius (middle), our multi-scale result (bottom)

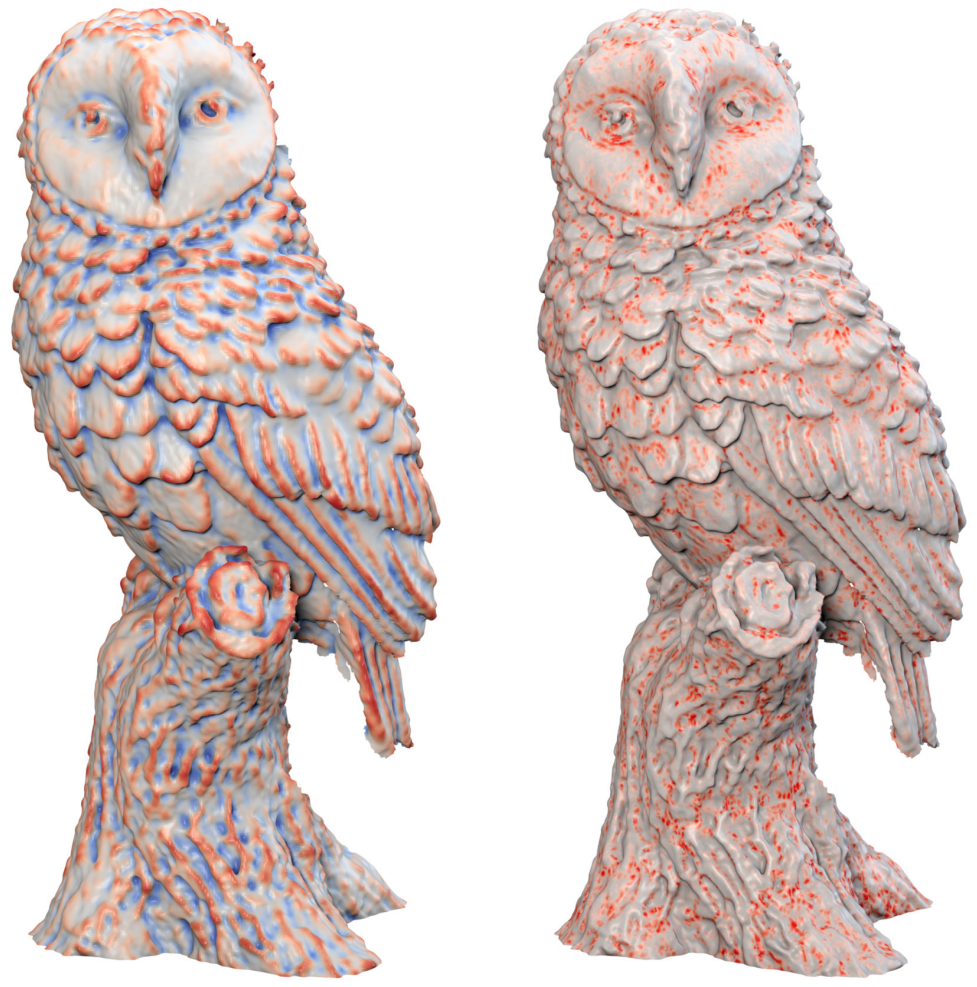

Figure 12: *Owl*: Our multi-scale curvature (left) is visually indistinguishable from the curvature field computed using a carefully hand-selected scale. The absolute differences (heavily amplified) are shown on the right.

## 5. Conclusion

In this paper, we revisited the problem of robust curvature estimation with a focus on multi-scale triangle meshes. Compared to previous approaches which ignore the presence of varying feature and noise scales, our algorithm is designed to take the local scale of the vertices into account. Our main contribution is the automatic computation of a mean curvature field, meaning that the radius of the ball is chosen for each vertex independently. We reviewed the performance and usefulness of our approach on several multi-scale datasets with respect to robust curvature estimation as well as adaptive mesh simplification using a density field as guidance. Even though our algorithm performs favorably on multi-scale data, we showed that this is not a requirement and that our algorithm also works on single-scale input data (Figure 12). In theory, the algorithm takes five optional parameters as input. In practice however, only the planar threshold $t_p$ and the initial radius factor $f_{initial}$ may be increased to achieve more smoothing if desired.

We see future work in optimizing our algorithm to reduce its runtime. In regions with very small triangles, many duplicate computations are performed when moving from one vertex to a neighboring. In a planar region that is highly tessellated, many CPU cycles are therefore wasted computing very similar curvatures. Furthermore, with a robust multi-scale curvature field at hand, other applications such as adaptive mesh smoothing seem promising.